\documentclass[aps,prl,twocolumn,superscriptaddress]{revtex4}

\usepackage{graphicx}
\usepackage{amsmath}
\usepackage{amssymb}
\usepackage{times}

\begin{document}


\title{Time Scales in Evolutionary Dynamics}


\author{Carlos P. Roca}
\homepage[]{http://www.gisc.es}

\author{Jos\'e A. Cuesta}
\homepage[]{http://www.gisc.es}

\affiliation{Grupo Interdisciplinar de Sistemas Complejos (GISC),
Departamento de Matem\'aticas, Universidad Carlos III de Madrid,
28911 Legan\'es, Madrid, Spain}

\author{Angel S\'anchez}
\homepage[]{http://www.gisc.es}

\affiliation{Grupo Interdisciplinar de Sistemas Complejos (GISC),
Departamento de Matem\'aticas, Universidad Carlos III de Madrid,
28911 Legan\'es, Madrid, Spain}

\affiliation{Instituto de Biocomputaci\'on y F\'isica de Sistemas
Complejos (BIFI), Universidad de Zaragoza,
50009 Zaragoza, Spain}


\date{\today}

\begin{abstract}

Evolutionary game theory has traditionally assumed that all
individuals in a population interact with each other between
reproduction events. We show that eliminating this restriction by
explicitly considering the time scales of interaction and selection
leads to dramatic changes in the outcome of evolution. Examples
include the selection of the inefficient strategy in the Harmony and
Stag-Hunt games, and the disappearance of the coexistence state in
the Snowdrift game. Our results hold for any population size and in
more general situations with additional factors influencing fitness.

\end{abstract}


\maketitle


Evolutionary game theory is the mathematical framework for modelling
evolution in biological, social and economical systems
\cite{Maynard-Smith:1982,Hofbauer:1998,Camerer:2003}, and is deeply
connected to dynamical systems theory and statistical mechanics
\cite{Szabo:2002,Traulsen:2004,Zimmermann:2004,Claussen:2005,
Santos:2005,Traulsen:2005,Fort:2005,Szabo:2005}. In the standard
setup of evolutionary game theory, strategies available for the game
are represented by a fraction of individuals in the population.
Individuals then interact according to the rules of the game, and the
so earned payoffs determine the frequencies of the next generation
(i.e., payoffs represent reproductive fitness). Customarily, most
evolutionary game studies make the additional assumption that
individuals play many times and with all other players before
reproduction takes place, so that payoffs, equivalently fitness, are
given by the mean distribution of types in the population. This is
also the situation for the so called round-robin tournament, in which
each individual plays once with every other. Both hypotheses,
common in biological evolution, implies that selection occurs much
more slowly than the interaction between individuals. Although recent
experimental studies show that this may not always be the case in
biology \cite{Hendry:1999,Hendry:2000,Yoshida:2003}, it is clear that
in cultural evolution or social learning the time scale of selection is
much closer to the time scale of interaction. The effects of this mixing
of scales cannot be disregarded \cite{Sanchez:2005}, and then it is
natural to ask about the consequences of the above assumption and the
effect of relaxing it. Though the main field of application of our
work is social and cultural evolution, we maintain the usual
language of evolutionary biology, to avoid introducing new
terminology.

In this Letter, we show that rapid selection affects evolutionary
dynamics in such a dramatic way that for some games it even changes
the stability of equilibria. In order to make explicit the relation
between selection and interaction time scales, we use discrete-time
dynamics. We follow Moran dynamics \cite{Moran:1962}, as this is the
proper way to describe evolution of discrete generations in the field of
population dynamics \cite{Roughgarden:1979}. Specifically, we choose
the frequency-dependent version of the Moran dynamics introduced by
\cite{Nowak:2004a}, which allows to consider an evolutionary game in
this dynamical context: $N$ individuals interact by playing a game
and reproduce by selecting one individual, with probability
proportional to the payoff, to duplicate and substitute a randomly
chosen individual. The payoff of every player is set to zero after
each reproduction event, and this two-step cycle is repeated until
the population eventually stabilizes. This stochastic dynamics is
discrete in both population and time, while keeping the population
size constant over time. Interestingly, this microscopic dynamics
leads to a difference equation that has been proposed as an adjusted
\cite{Maynard-Smith:1982} or discrete-time \cite{Hofbauer:1998}
analogous of the replicator equation, widely used in evolutionary
game theory (see \cite{Traulsen:2005} for a recent, detailed
discussion of this issue). Additionally, we note that for social
applications, reproduction may be also interpreted as a learning
process, in which individuals do not die but instead change the way
they behave or their strategies.

Time scales enter the dynamics through the interaction step,
affecting the way fitness is obtained. We introduce a new interaction
scheme, by allowing an integer number $s$ of randomly chosen pairs of
individuals to play consecutively the game, between reproduction
events. Thus, $s$ equals the ratio between selection and
interaction time scales. This is the crucial parameter in our model.
The limit value of $s=1$ means that both time scales are equal;
greater finite values, $s>1$, correspond to the selection time scale
being slower than the interaction time scale, and the limit value of
$s\to\infty$ recovers the round-robin procedure. In fact, the
equivalence of the limit $s\to\infty$ to the round-robin scheme
points to the latter being a form of 'mean-field' theory, in which
individuals reproduce so slowly that it makes sense to replace
pairwise interactions by the interaction with the 'average player'.

As for the games, we will consider the important case of symmetric
$2 \times 2$ games, in which the payoffs are given by the following matrix
\begin{equation} \label{eq:2by2game}
\begin{array}{ccc}
 & \mbox{ }\, 1 & \!\!\!\!\!\! 2 \\
\begin{array}{c} 1 \\ 2 \end{array} & \left(\begin{array}{c} a \\ c
\end{array}\right. & \left.\begin{array}{c} b \\ d
\end{array}\right), \end{array} \end{equation} whose rows give the
payoff obtained by each strategy when confronted with the other or
itself, and $a,b,c,d>0$. Let $n$ be the number of individuals using
strategy 1, also referred as type 1 individuals. After each
reproduction event $n$ may stay the same, increase by one, or
decrease by one. Considering the definition of the dynamics, the
corresponding transition probabilities will depend on the fitness
earned by each type during the interaction step and on their
frequencies. As both quantities will depend ultimately on $n$, we
have a Markov process with a tridiagonal transition matrix (i.e., a
birth-death process \cite{Karlin:1975}) whose non-zero coefficients
are \begin{equation}\begin{split}
P_{n,n-1} &= \frac{n}{N} E\left(\frac{F_2}{F_1+F_2} \:\bigg|\: n \right) \\
P_{n,n+1} &= \frac{N-n}{N} E\left(\frac{F_1}{F_1+F_2} \:\bigg|\: n
\right) \end{split} \label{eq:tranprob} \end{equation} and $P_{n,n} =
1 - P_{n,n-1} - P_{n,n+1}$. $F_i$ is the payoff obtained by all
players of type $i$, and $E( \cdot | n )$ denotes the expected value
conditioned to a population of $n$ individuals of type 1.

We stress that the parameter $s$ enters through the expected values
of the relative fitness of each type (\ref{eq:tranprob}). Indeed, if
we restrict ourselves to the limit $s\to\infty$,
these expected values are given directly by the pairing probabilities
and the payoffs corresponding to each pair
\begin{eqnarray}
 && E \left(\left.\frac{F_1}{F_1+F_2} \:\right|\: n \right) =  \\
 &&\frac{n(n-1)a + n(N-n)b}{n(n-1)a + n(N-n)(b+c) + (N-n)(N-n-1)d} \nonumber
\end{eqnarray} as would be obtained by the round-robin scheme.
However, as we will see below, finite values of $s$ often lead to
results completely different from this special case.

The solution to the birth-death process we have just described
can be obtained in a
standard manner \cite{Karlin:1975}. Denoting by $p_n$ the
fixation probability of type 1 (i.e. the probability of ending up in a
population with all individuals of type 1) when starting from a
population with $n$ players of this type, we have
\begin{equation}
p_n = P_{n,n-1}p_{n-1} + P_{n,n}p_n + P_{n,n+1}p_{n+1},
\end{equation}
with $p_0=0$ and $p_N=1$. The solution to this
equation is given by
\begin{equation}
\label{eq:pn}
p_n = Q_n/Q_N, \quad Q_n =
\displaystyle 1 + \sum_{j=1}^{n-1}\prod_{i=1}^j
\frac{P_{i,i-1}}{P_{i,i+1}}, \quad n>1
\end{equation}
with $Q_1=1$.
As stated above, the interesting case arises for
finite values of the parameter $s$. For general $s$,
a straightforward combinatorial analysis of
all the possible sequences of $s$ pairings leads to
\begin{widetext}
\begin{equation}
\label{eq:expvalue}
E\left(\frac{F_1}{F_1+F_2} \:\Big|\: n \right) = \sum_{i=0}^s \sum_{j=0}^{s-i} \left[ 2^{s-i-j}
\frac {s!n^{s-j}(n-i)^i(N-n)^{s-i}(N-n-1)^j} {i!j!(s-i-j)!(N(N-1))^s }
 \frac {2ai + b(s-i-j)} {2ai + 2dj + (b+c)(s-i-j)} \right].
\end{equation}
\end{widetext}
This lengthy combinatorial expression reduces, in the limit case $s=1$
of extremely rapid selection, to
\begin{equation}\begin{split}
P_{n,n-1} &= \frac{n(N-n)}{N(N-1)} \left( 1
+ \frac{c-b}{c+b}\frac{n}{N} - \frac{1}{N} \right) \\
P_{n,n+1} &= \frac{n(N-n)}{N(N-1)} \left( \frac{2b}{b+c} +
\frac{c-b}{c+b}\frac{n}{N} - \frac{1}{N} \right). \end{split}
\label{eq:tranprob-s1} \end{equation} The above equations are the
first hint of the effect of time scales. Indeed, by noting that, for
this extreme case, only the coefficients of the skew diagonal of
(\ref{eq:2by2game}) appear in (\ref{eq:tranprob-s1}) we reach the
surprising conclusion that if the time scale of selection equals that
of interaction, the evolutionary outcome of any game will be
determined solely by the performance of each strategy when confronted
with the other, and independently of the results when dealing with
itself. However, as we will now see, there are another non-trivial,
important differences.

To make our study as general as possible, we have analyzed all twelve
non-equivalent symmetric $2 \times 2$ games \cite{Rapoport:1966}.
These games can be further classified into three categories,
according to their Nash equilibria and their dynamical behavior
under the replicator dynamics with round-robin interaction:

I.\ There are six games with $a>c$ and $b>d$, or $a<c$ and $b<d$.
They have a unique Nash equilibrium, corresponding to the dominant
pure strategy. This equilibrium is the global attractor of the
round-robin replicator dynamics.

II.\ There are three games with $a>c$ and $b<d$. They have several
Nash equilibria, one of them with a mixed strategy. With the
round-robin replicator dynamics, this mixed strategy equilibrium is
an unstable point, which acts as the boundary between the basins of
attraction of the two pure strategies, which are the attractors.

III.\ The remaining three games have $a<c$ and $b>c$. They have
several Nash equilibria, one of them with a mixed strategy. This
mixed strategy equilibrium is the global attractor of the round-robin
replicator dynamics. The two pure strategies are unstable in this
case.

\begin{figure} \includegraphics[width=7.2cm]{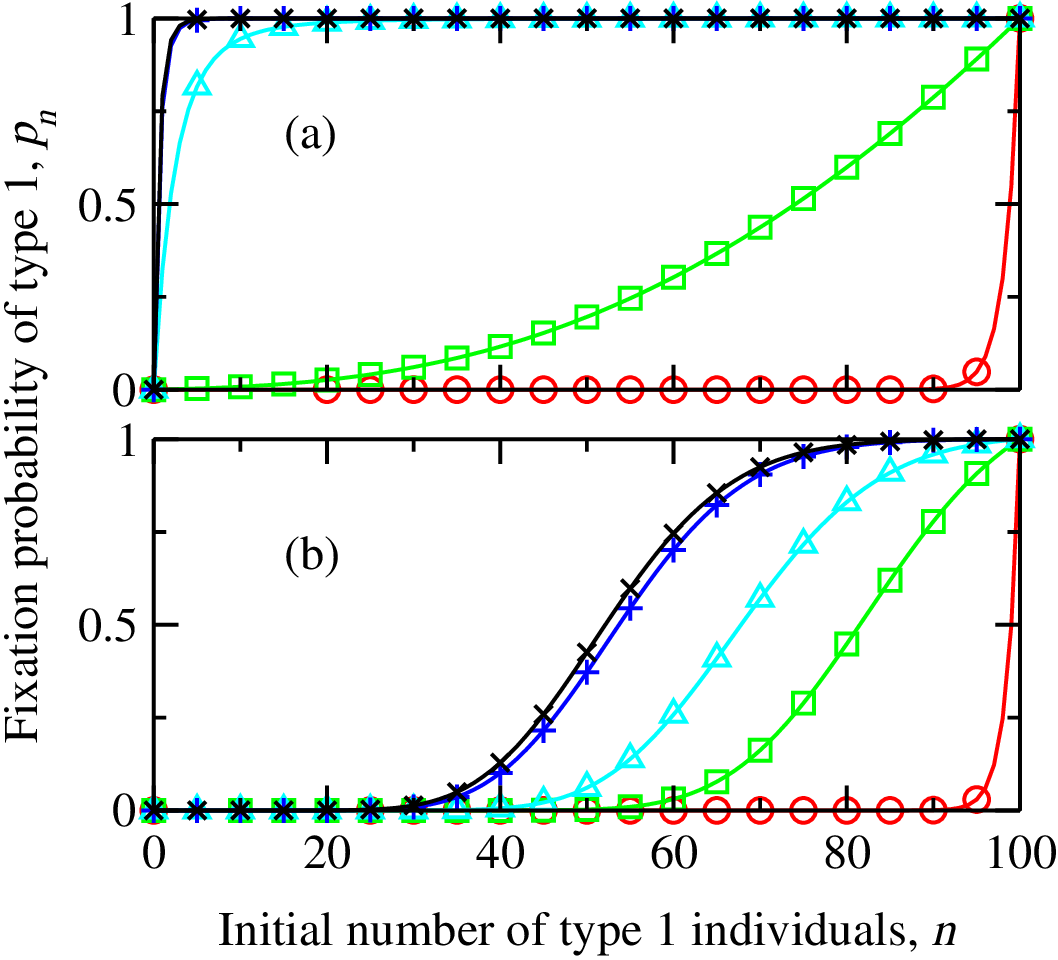}
\caption{\label{fig:1}Fixation probabilities in the games (a) Harmony
($a=1$, $b=0.25$, $c=0.75$, $d=0.01$) and (b) Stag-Hunt ($a=1$,
$b=0.01$, $c=0.8$, $d=0.2$), for $s$ equal to 1 ($\circ$), 5
($\Box$), 10 ($\triangle$), 100 ($+$), or $\to\infty$ ($\times$).
Note that, in figure (a), curves overlap for $s=10, 100$ and
$\to\infty$. Population size $N=100$.} \end{figure}

Let us first consider an example of class I, namely the Harmony game
\cite{Licht:1999} ($a=1$, $b=0.25$, $c=0.75$, $d=0.01$). This is a
no-conflict game, in which all players obtain the maximum payoff by
following strategy 1. As Fig.\ \ref{fig:1}(a) shows, this is the
result for large values of $s$, with a fixation probability $p_n
\approx 1$ for almost all $n$. On the other hand, Fig.\
\ref{fig:1}(a) also shows that, for small $s$, strategy 2, i.e., the
inefficient (in the sense of lowest payoff) one, is selected by the
dynamics, unless starting from initial conditions with almost all
individuals of type 1.

For class II, a good paradigm is the Stag-Hunt game
\cite{Skirms:2003} ($a=1$, $b=0.01$, $c=0.8$, $d=0.2$),
which is a coordination game: Strategy 1 maximizes the
mutual benefit, whereas strategy 2
minimizes the risk of loss, and the conflict results from
having to choose between these two options. As Fig.\
\ref{fig:1}(b) reveals, the round-robin result is obtained for large
$s$: both strategies are attractors, with the basin boundary located
at the frequency corresponding to the mixed strategy equilibrium,
i.e. $x = (d-b)/(a-c+d-b) \approx 0.49$. However, for small values of
$s$ this boundary shifts to greater frequency values, thus reflecting
an advantage of strategy 2. In the extreme $s = 1$ case this strategy
becomes the unique attractor.

It is interesting to note that Fig.\ \ref{fig:1} shows that
there is not a general crossover at $s \approx N$. In the Harmony
game, the round-robin regime is mostly reached for $s \simeq 10\ll N$,
whereas in the Stag-Hunt game this does not happen until $s \simeq
100 = N$.

\begin{figure} \includegraphics[width=7.64cm]{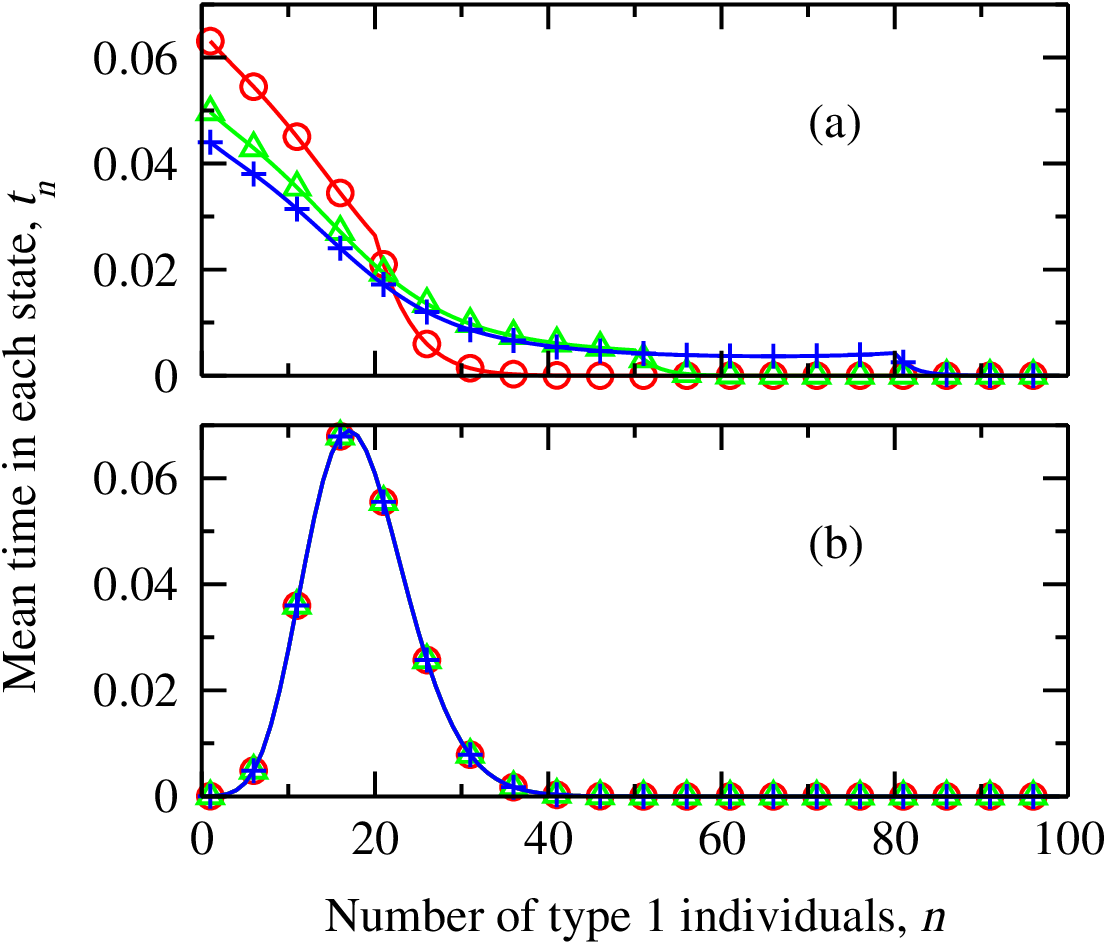}
\caption{\label{fig:2}Mean time before fixation in the Snowdrift game
($a=1$, $b=0.2$, $c=1.8$, $d=0.01$), for $s$ equal to 5 (a) and 100
(b). Initial values of $n$ equal to 20 ($\circ$), 50 ($\triangle$)
and 80 ($+$). Note that curves in (b) overlap. Population size
$N=100$.} \end{figure}

Finally, let us consider the Snowdrift game \cite{Sudgen:2004}
($a=1$, $b=0.2$, $c=1.8$, $d=0.01$) as an example of class III. This
is also a dilemma game, as each player has to choose between strategy
1, which maximizes the population gain, and strategy 2, which gives
individuals the maximum payoff by exploiting the opponent. With
round-robin dynamics both strategies coexists in the long run, with
frequencies corresponding to the mixed strategy equilibrium. However,
our dynamics can never maintain coexistence indefinitely, because by
construction one of the absorbing states (all players of type 1 or
all of type 2) will be reached sooner or later with probability 1.
Nonetheless, it is possible to study the duration of metastable
states by using the mean time in each population state before
absorbtion, $t_n$ \cite{Karlin:1975}. Figure \ref{fig:2} shows the
results for two values of $s$ and a broad range of initial
conditions. For $s$ large ($s=100$), the population stays for a long
time near the value corresponding to the mixed strategy equilibrium
$x = (d-b)/(a-c+d-b) \approx 0.19$, independently of the initial
number of type 1 individuals. A smaller value of $s=10$ (not shown)
induces a shift of the metastable equilibrium to smaller values of
$n$, again almost independently of the initial conditions. Finally,
for an even smaller value of $s$ ($s=5$), there is no metastable
equilibrium, but a fluctuation towards the $x=0$ absorbing state,
which clearly depends on the initial conditions.

Having given examples of all three classes, we will summarize the
rest of our study by saying that the remaining $2 \times 2$ games
behave in a similar way, with rapid selection (small $s$) favoring in
all cases the type that has the greatest coefficient in the skew
diagonal of the payoff matrix. For the remaining five games of class
I this results in a reinforcement of the dominant strategy (the
Prisoner's Dilemma \cite{Axelrod:1981} being a prominent example).
The other two games of class II exhibit once again a displacement of
the basins of attraction, whereas the other two class III games
display the suppression of the coexistence state in favor of one of
the strategies. We thus see that rapid selection leads very generally
to outcomes entirely different from those of round-robin dynamics.

It is important to realize that our results do not change
qualitatively with the system size. Considering for instance the
Stag-Hunt game, the change in the basins of attraction is practically
independent of the population size. The main effect of working with
larger sizes is a steeper transition between the basins of
attraction. Indeed, due to the inherent stochasticity of finite
population sizes, smaller populations have a more blurred basin
boundary, with points in each basin having an increasing non-zero
probability of reaching the other basin \cite{Cabrales:2000}. Our
results for all other symmetric $2 \times 2$ games are equally
robust. In fact, for very rapid selection, $s=1$, the limit $N \to
\infty$ of the transition probabilities, Eq. (\ref{eq:tranprob-s1}),
shows that they depend only on the frequencies of both types.

\begin{figure} \includegraphics[width=7.2cm]{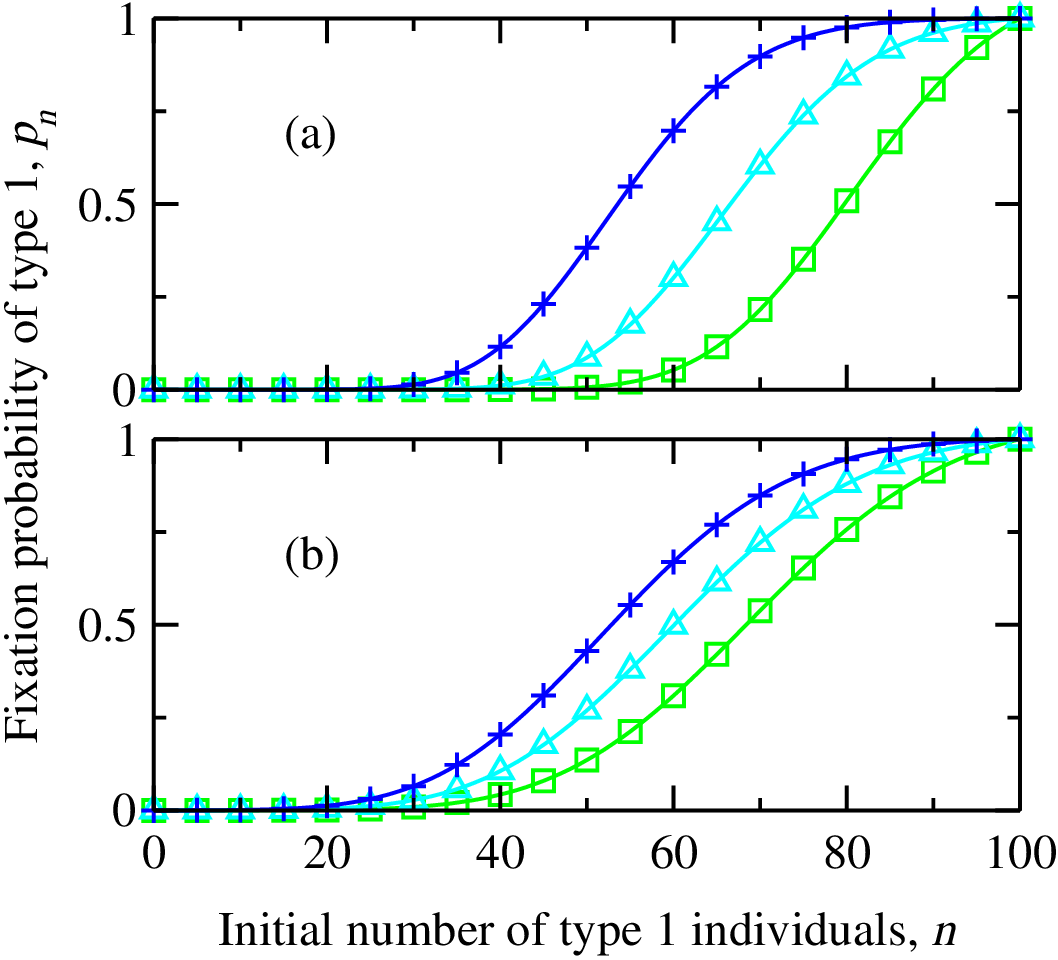}
\caption{\label{fig:3} Fixation probability in the Stag-Hunt game
($a=1$, $b=0.01$, $c=0.8$, $d=0.2$) with a background of fitness
$f_b$ equal to 0.1 (a) and 1 (b). Values of $s$: 5 ($\Box$), 10
($\triangle$) 100 ($+$). Population size $N=100$.} \end{figure}

It could be argued that in our model only $s$ pairs of individuals
play in each round, resulting in a very small effective population,
this being the fundamental cause of the reported results. To probe
into this issue, we have introduced a background of fitness
\cite{Claussen:2005,Nowak:2004a}, so that every player has an
intrinsic probability of being selected, regardless of the outcome of
the game, and thus guaranteeing a population of $N$ players. Indeed,
in most applications, agents interact through more than one type of
game and there are external contributions to fitness(environmental factors,
fashions or media influence in a social context, etc.). Let $f_b$ 
be the normalized
fitness background, so that each individual has a background of
fitness $s f_b / N$ before selection takes place; $f_b=1$ means that
the overall fitness coming from the game and from the background are
approximately equal, for every value of $s$ and $N$. Figure
\ref{fig:3} shows the results for the Stag-Hunt game. A small fitness
background of $f_b=0.1$ gives fixation probabilities very similar to
those with $f_b=0$ (Fig. \ref{fig:2}(b)). For larger values, $f_b=1$,
the displacement of the basin boundary is smaller, but still
perfectly noticeable. And a very large fitness background, $f_b
\gtrsim 10$ (not shown), drives the dynamics to random selection for
every value of $s$, because in this case the influence of the game is
almost negligible. Again, for the remaining symmetric $2 \times 2$
games, our conclusions remain valid as well in the presence of a
background of fitness. Consequently, our results are not merely due
to a finite size effect of a small effective population of players.

In summary, we have proven that considering independent interaction
and selection time scales leads to highly non-trivial,
counter-intuitive results. We have demonstrated the generality of
this conclusion by considering all symmetric $2 \times 2$ games and
showing that rapid selection may lead to changes of the
asymptotically selected equilibria, to changes of the basins of
attraction of equilibria, or to suppression of long-lived metastable
equilibria. This result has major implications for applying
evolutionary game theory to model a specific problem, as the
assumption of slow selection and consequently of round-robin dynamics
may or may not be correct. Indeed, as the example in
\cite{Sanchez:2005} shows, rapid selection may lead to the
understanding of problems where Darwinian, individual evolution was
thought not to play a role because round-robin dynamics was used. We
envisage that successful modelling in rapidly changing environments,
such as social or (sub-)culture dynamics, will need a careful
consideration of the involved time scales along the lines discussed
here.

We thank R.\ Toral for a critical reading of the manuscript. This
work is supported by MEC (Spain) under grants
BFM2003-0180, BFM2003-07749-C05-01,
FIS2004-1001 and NAN2004-9087-C03-03 and by Comunidad de Madrid
(Spain) under grants
UC3M-FI-05-007, SIMUMAT-CM and MOSSNOHO-CM.

\bibliography{prl}

\begin{thebibliography}{25}
\expandafter\ifx\csname natexlab\endcsname\relax\def\natexlab#1{#1}\fi
\expandafter\ifx\csname bibnamefont\endcsname\relax
  \def\bibnamefont#1{#1}\fi
\expandafter\ifx\csname bibfnamefont\endcsname\relax
  \def\bibfnamefont#1{#1}\fi
\expandafter\ifx\csname citenamefont\endcsname\relax
  \def\citenamefont#1{#1}\fi
\expandafter\ifx\csname url\endcsname\relax
  \def\url#1{\texttt{#1}}\fi
\expandafter\ifx\csname urlprefix\endcsname\relax\def\urlprefix{URL }\fi
\providecommand{\bibinfo}[2]{#2}
\providecommand{\eprint}[2][]{\url{#2}}

\bibitem[{\citenamefont{Maynard~Smith}(1982)}]{Maynard-Smith:1982}
\bibinfo{author}{\bibfnamefont{J.}~\bibnamefont{Maynard~Smith}},
  \emph{\bibinfo{title}{Evolution and the Theory of Games}}
  (\bibinfo{publisher}{Cambridge University Press}, \bibinfo{year}{1982}).

\bibitem[{\citenamefont{Hofbauer and Sigmund}(1998)}]{Hofbauer:1998}
\bibinfo{author}{\bibfnamefont{J.}~\bibnamefont{Hofbauer}} \bibnamefont{and}
  \bibinfo{author}{\bibfnamefont{K.}~\bibnamefont{Sigmund}},
  \emph{\bibinfo{title}{Evolutionary Games and Population Dynamics}}
  (\bibinfo{publisher}{Cambridge University Press}, \bibinfo{year}{1998}).

\bibitem[{\citenamefont{Camerer}(2003)}]{Camerer:2003}
\bibinfo{author}{\bibfnamefont{C.~F.} \bibnamefont{Camerer}},
  \emph{\bibinfo{title}{Behavioral Game Theory}} (\bibinfo{publisher}{Princeton
  University Press}, \bibinfo{year}{2003}).

\bibitem[{\citenamefont{Szab{\'o} and Hauert}(2002)}]{Szabo:2002}
\bibinfo{author}{\bibfnamefont{G.}~\bibnamefont{Szab{\'o}}} \bibnamefont{and}
  \bibinfo{author}{\bibfnamefont{C.}~\bibnamefont{Hauert}},
  \bibinfo{journal}{Phys. Rev. Lett.} \textbf{\bibinfo{volume}{89}},
  \bibinfo{pages}{118101} (\bibinfo{year}{2002}).

\bibitem[{\citenamefont{Traulsen et~al.}(2004)\citenamefont{Traulsen, R{\"o}hl,
  and Schuster}}]{Traulsen:2004}
\bibinfo{author}{\bibfnamefont{A.}~\bibnamefont{Traulsen}},
  \bibinfo{author}{\bibfnamefont{T.}~\bibnamefont{R{\"o}hl}}, \bibnamefont{and}
  \bibinfo{author}{\bibfnamefont{H.~G.} \bibnamefont{Schuster}},
  \bibinfo{journal}{Phys. Rev. Lett.} \textbf{\bibinfo{volume}{93}},
  \bibinfo{pages}{28701} (\bibinfo{year}{2004}).

\bibitem[{\citenamefont{Zimmermann et~al.}(2004)\citenamefont{Zimmermann,
  Egu{\'\i}luz, and {San~Miguel}}}]{Zimmermann:2004}
\bibinfo{author}{\bibfnamefont{M.~G.} \bibnamefont{Zimmermann}},
  \bibinfo{author}{\bibfnamefont{V.~M.} \bibnamefont{Egu{\'\i}luz}},
  \bibnamefont{and}
  \bibinfo{author}{\bibfnamefont{M.}~\bibnamefont{{San~Miguel}}},
  \bibinfo{journal}{Phys. Rev. E} \textbf{\bibinfo{volume}{69}},
  \bibinfo{pages}{65102} (\bibinfo{year}{2004}).

\bibitem[{\citenamefont{Claussen and Traulsen}(2005)}]{Claussen:2005}
\bibinfo{author}{\bibfnamefont{J.~C.} \bibnamefont{Claussen}} \bibnamefont{and}
  \bibinfo{author}{\bibfnamefont{A.}~\bibnamefont{Traulsen}},
  \bibinfo{journal}{Phys. Rev. E} \textbf{\bibinfo{volume}{71}},
  \bibinfo{pages}{25101} (\bibinfo{year}{2005}).

\bibitem[{\citenamefont{Santos and Pacheco}(2005)}]{Santos:2005}
\bibinfo{author}{\bibfnamefont{F.~C.} \bibnamefont{Santos}} \bibnamefont{and}
  \bibinfo{author}{\bibfnamefont{J.~M.} \bibnamefont{Pacheco}},
  \bibinfo{journal}{Phys. Rev. Lett.} \textbf{\bibinfo{volume}{95}},
  \bibinfo{pages}{98104} (\bibinfo{year}{2005}).

\bibitem[{\citenamefont{Traulsen et~al.}(2005)\citenamefont{Traulsen, Claussen,
  and Hauert}}]{Traulsen:2005}
\bibinfo{author}{\bibfnamefont{A.}~\bibnamefont{Traulsen}},
  \bibinfo{author}{\bibfnamefont{J.~C.} \bibnamefont{Claussen}},
  \bibnamefont{and} \bibinfo{author}{\bibfnamefont{C.}~\bibnamefont{Hauert}},
  \bibinfo{journal}{Phys. Rev. Lett.} \textbf{\bibinfo{volume}{95}},
  \bibinfo{pages}{238701} (\bibinfo{year}{2005}).

\bibitem[{\citenamefont{Ariosa and Fort}(2005)}]{Fort:2005}
\bibinfo{author}{\bibfnamefont{D.}~\bibnamefont{Ariosa}} \bibnamefont{and}
  \bibinfo{author}{\bibfnamefont{H.}~\bibnamefont{Fort}},
  \bibinfo{journal}{Phys. Rev. E} \textbf{\bibinfo{volume}{71}},
  \bibinfo{pages}{16132} (\bibinfo{year}{2005}).

\bibitem[{\citenamefont{Hauert and Szab{\'o}}(2005)}]{Szabo:2005}
\bibinfo{author}{\bibfnamefont{C.}~\bibnamefont{Hauert}} \bibnamefont{and}
  \bibinfo{author}{\bibfnamefont{G.}~\bibnamefont{Szab{\'o}}},
  \bibinfo{journal}{Am. J. Phys.} \textbf{\bibinfo{volume}{73}},
  \bibinfo{pages}{405} (\bibinfo{year}{2005}).

\bibitem[{\citenamefont{Hendry and Kinnison}(1999)}]{Hendry:1999}
\bibinfo{author}{\bibfnamefont{A.~P.} \bibnamefont{Hendry}} \bibnamefont{and}
  \bibinfo{author}{\bibfnamefont{M.~T.} \bibnamefont{Kinnison}},
  \bibinfo{journal}{Evolution} \textbf{\bibinfo{volume}{53}},
  \bibinfo{pages}{1637} (\bibinfo{year}{1999}).

\bibitem[{\citenamefont{Hendry et~al.}(2000)\citenamefont{Hendry, Wenburg,
  Bentzen, Volk, and Quinn}}]{Hendry:2000}
\bibinfo{author}{\bibfnamefont{A.~P.} \bibnamefont{Hendry}},
  \bibinfo{author}{\bibfnamefont{J.~K.} \bibnamefont{Wenburg}},
  \bibinfo{author}{\bibfnamefont{P.}~\bibnamefont{Bentzen}},
  \bibinfo{author}{\bibfnamefont{E.~C.} \bibnamefont{Volk}}, \bibnamefont{and}
  \bibinfo{author}{\bibfnamefont{T.~P.} \bibnamefont{Quinn}},
  \bibinfo{journal}{Science} \textbf{\bibinfo{volume}{290}},
  \bibinfo{pages}{516} (\bibinfo{year}{2000}).

\bibitem[{\citenamefont{Yoshida et~al.}(2003)\citenamefont{Yoshida, Jones,
  Ellner, Fussmann, and Jr.}}]{Yoshida:2003}
\bibinfo{author}{\bibfnamefont{T.}~\bibnamefont{Yoshida}},
  \bibinfo{author}{\bibfnamefont{L.~E.} \bibnamefont{Jones}},
  \bibinfo{author}{\bibfnamefont{S.~P.} \bibnamefont{Ellner}},
  \bibinfo{author}{\bibfnamefont{G.~F.} \bibnamefont{Fussmann}},
  \bibnamefont{and} \bibinfo{author}{\bibfnamefont{N.~G.~H.}
  \bibnamefont{Jr.}}, \bibinfo{journal}{Nature} \textbf{\bibinfo{volume}{424}},
  \bibinfo{pages}{303} (\bibinfo{year}{2003}).

\bibitem[{\citenamefont{S\'anchez and Cuesta}(2005)}]{Sanchez:2005}
\bibinfo{author}{\bibfnamefont{A.}~\bibnamefont{S\'anchez}} \bibnamefont{and}
  \bibinfo{author}{\bibfnamefont{J.~A.} \bibnamefont{Cuesta}},
  \bibinfo{journal}{J. Theor. Biol.} \textbf{\bibinfo{volume}{235}},
  \bibinfo{pages}{233} (\bibinfo{year}{2005}).

\bibitem[{\citenamefont{Moran}(1962)}]{Moran:1962}
\bibinfo{author}{\bibfnamefont{P.~A.~P.} \bibnamefont{Moran}},
  \emph{\bibinfo{title}{The Statistical Processes of Evolutionary Theory}}
  (\bibinfo{publisher}{Clarendon Press}, \bibinfo{address}{Oxford},
  \bibinfo{year}{1962}).

\bibitem[{\citenamefont{Roughgarden}(1979)}]{Roughgarden:1979}
\bibinfo{author}{\bibfnamefont{J.}~\bibnamefont{Roughgarden}},
  \emph{\bibinfo{title}{Theory of Population Genetics and Evolutionary Ecology:
  An Introduction}} (\bibinfo{publisher}{MacMillan Publishing Co.},
  \bibinfo{address}{New York}, \bibinfo{year}{1979}).

\bibitem[{\citenamefont{Nowak et~al.}(2004)\citenamefont{Nowak, Sasaki, Taylor,
  and Fudenberg}}]{Nowak:2004a}
\bibinfo{author}{\bibfnamefont{M.~A.} \bibnamefont{Nowak}},
  \bibinfo{author}{\bibfnamefont{A.}~\bibnamefont{Sasaki}},
  \bibinfo{author}{\bibfnamefont{C.}~\bibnamefont{Taylor}}, \bibnamefont{and}
  \bibinfo{author}{\bibfnamefont{D.}~\bibnamefont{Fudenberg}},
  \bibinfo{journal}{Nature} \textbf{\bibinfo{volume}{428}},
  \bibinfo{pages}{646} (\bibinfo{year}{2004}).

\bibitem[{\citenamefont{Karlin and Taylor}(1975)}]{Karlin:1975}
\bibinfo{author}{\bibfnamefont{S.}~\bibnamefont{Karlin}} \bibnamefont{and}
  \bibinfo{author}{\bibfnamefont{H.~M.} \bibnamefont{Taylor}},
  \emph{\bibinfo{title}{A First Course in Stochastic Processes}}
  (\bibinfo{publisher}{Academic Press}, \bibinfo{address}{New York},
  \bibinfo{year}{1975}), \bibinfo{edition}{2nd} ed.

\bibitem[{\citenamefont{Rapoport and Guyer}(1966)}]{Rapoport:1966}
\bibinfo{author}{\bibfnamefont{A.}~\bibnamefont{Rapoport}} \bibnamefont{and}
  \bibinfo{author}{\bibfnamefont{M.}~\bibnamefont{Guyer}},
  \bibinfo{journal}{General Systems} \textbf{\bibinfo{volume}{11}},
  \bibinfo{pages}{203} (\bibinfo{year}{1966}).

\bibitem[{\citenamefont{Licht}(1999)}]{Licht:1999}
\bibinfo{author}{\bibfnamefont{A.~N.} \bibnamefont{Licht}},
  \bibinfo{journal}{Yale J. Int. Law} \textbf{\bibinfo{volume}{24}},
  \bibinfo{pages}{61} (\bibinfo{year}{1999}).

\bibitem[{\citenamefont{Skyrms}(2003)}]{Skirms:2003}
\bibinfo{author}{\bibfnamefont{B.}~\bibnamefont{Skyrms}},
  \emph{\bibinfo{title}{The Stag Hunt and the Evolution of Social Structure}}
  (\bibinfo{publisher}{Cambridge University Press}, \bibinfo{year}{2003}).

\bibitem[{\citenamefont{Sugden}(2004)}]{Sudgen:2004}
\bibinfo{author}{\bibfnamefont{R.}~\bibnamefont{Sugden}},
  \emph{\bibinfo{title}{Economics of Rights, Co-operation and Welfare}}
  (\bibinfo{publisher}{Palgrave Macmillan}, \bibinfo{address}{Hampshire},
  \bibinfo{year}{2004}), \bibinfo{edition}{2nd} ed.

\bibitem[{\citenamefont{Axelrod and Hamilton}(1981)}]{Axelrod:1981}
\bibinfo{author}{\bibfnamefont{R.}~\bibnamefont{Axelrod}} \bibnamefont{and}
  \bibinfo{author}{\bibfnamefont{W.~D.} \bibnamefont{Hamilton}},
  \bibinfo{journal}{Science} \textbf{\bibinfo{volume}{211}},
  \bibinfo{pages}{1390} (\bibinfo{year}{1981}).

\bibitem[{\citenamefont{Cabrales}(2000)}]{Cabrales:2000}
\bibinfo{author}{\bibfnamefont{A.}~\bibnamefont{Cabrales}},
  \bibinfo{journal}{Int. Econ. Rev.} \textbf{\bibinfo{volume}{41}},
  \bibinfo{pages}{451} (\bibinfo{year}{2000}).

\end{thebibliography}

\end{document}